\documentclass[a4,12pt]{article}
\usepackage{epsfig}

\def\tstrut#1#2{\hbox{\vrule height #1 pt depth #2 pt width 0 pt}}

\def\cL{{\cal L}}

\def\gm{\gamma}

\def\ovl{\overline}

\def\bra{\langle}

\def\ket{\rangle}
\def\dspl{\displaystyle}
\def\lrlap#1{\hbox to 0pt{\hss#1\hss}}

\def\beq{\begin{equation}}
\def\eeq{\end{equation}}
\def\beqa{\begin{eqnarray}}
\def\eeqa{\end{eqnarray}}
\def\bec{\begin{center}}
\def\enc{\end{center}}

\begin{document}

\title{CP Violation in B Decays}

\author{Hitoshi Yamamoto\\
The University of Hawaii \\
 2505 Correa Rd, Honolulu, HI 96822, USA
\\E-mail: hitoshi@phys.hawaii.edu
}

\maketitle

\abstract{
We review the physics of $CP$ violation in $B$ decays.
After introducing the $CKM$ matrix and how it causes $CP$ violation,
we cover three types of $CP$ violation that can occur in
B decays: $CP$ violation in mixing, $CP$ violation by
mixing-decay interference, and $CP$ violation in decay.}

\section{$CP$ Violation and the CKM Matrix}
\subsection{$CP$ Transformation of the quark-$W$ interaction}
General left-handed quark-$W$ interaction can be written in the interaction
picture as (for 3 generations of quarks)
 \beq
  L_{\rm int}(t) = \int d^3x \bigg(\cL_{qW}(x) + \cL_{qW}^\dag(x)\bigg)
 \label{eq:lint}
\eeq
which is the space-integral of the Lagrangian density given by
\beq
  \cL_{qW}(x) = {g\over\sqrt8} \sum_{i,\,j = 1,3} V_{i\,j}\;
    \bar U_i(x) \;\gm_\mu (1 - \gm_5)D_j(x)\;
    W^\mu(x)
 \label{eq:lqw}
\eeq
where $U_i$ and $D_i$ are the up-type and down-type quark fields
\beq
    U_i(x) \equiv \pmatrix{ u(x) \cr c(x) \cr t(x) }\,,\quad
    D_j(x) \equiv \pmatrix{ d(x) \cr s(x) \cr b(x) }\qquad
    (x \equiv (t,\vec x))
\eeq
and the coupling of the $V-A$ quark currents to $W$ is given by the
complex parameters $V_{ij}$ forming a $3\times3$ matrix
\beq
   V =
  \pmatrix{ V_{ud} & V_{us} & V_{ub} \cr
            V_{cd} & V_{cs} & V_{cb} \cr
            V_{td} & V_{ts} & V_{tb} }\quad 
\eeq
called the Cabbibo-Kobayashi-Masukawa ($CKM$) matrix.

The $CP$ transformation exchanges particle $n$ and its antiparticle
$\bar n$, flips the momentum, and keeps the spin $z$ component $\sigma$ unchanged.
In terms of creation operators,
\beq
      (CP) a^\dag_{n,\vec p,\sigma} (CP)^\dag
     \equiv \eta_n\, a^\dag_{\bar n,-\vec p,\sigma}
 \label{eq:cpadag}
\eeq
where $\eta_n$ is an arbitrary phase factor that in general can depend on
particle type (except that those of a particle and its antiparticle are
related by $\eta_{\bar n} = (-)^{2J} \eta^*_n$ where
$J$ is the spin of the particle) 
that in essence defines the $CP$ operator in the Hilbert
space. Any choice of the $CP$ phases gives a legitimate $CP$ operator. For some
choices, however, a given interaction Lagrangian may commute with the 
$CP$ operator, and if such choice can be made, then
the processes caused by the interaction is invariant under $CP$.

Quark and W fields are made of creation and annihilation operators,
and thus the transformation property (\ref{eq:cpadag}) leads to
those of fields. Then, a straightforward algebra shows that
the quark-W interaction (\ref{eq:lqw}) transforms as (setting
the irrelevant phase of $W$, $\eta_W$, to be unity)
\beq
  (CP) \cL_{qW}(CP)^\dag = 
 {g\over\sqrt8}\sum_{i,j=1,3} \eta_{U_i}\eta_{D_j}^* V_{ij}\;
  (\bar U_i\gm^\mu(1-\gm_5)D_j W_\mu)^\dag\,,
\eeq
where the space-time argument $x$ on the left changed to $x'=(t,-\vec x)$
on the right which has no significance when integrated over space.
Then, if one can choose the phases such that
\beq
     \eta_{U_i}\eta_{D_j}^* V_{ij} = V_{ij}^*\,,
\eeq
then we have $(CP) \cL_{qW} (CP)^\dag = \cL_{qW}^\dag$ and the two
terms in (\ref{eq:lint}) simply swaps keeing the interaction Lagrangian
invariant under $CP$. Given that $\eta_{U_i}$ and $\eta_{D_j}$ are
arbitrary phases associated with each quark, 
the condition above is equivalent to being able to
rotate quark phases to make all $V_{ij}$ real without changing
$\cL_{qW}$. 
We can always make 5 of $V_{ij}$ real since
there are 6 quarks which have 5 relative phases.

\subsection{Unitarity triangle}
So far, we dealt with a completely general $3\times 3$ matrix $V$.
In the standard model, the CKM matrix is written as $V = S^{u\dag}S^d$
where $S^{u(d)}$ is the unitary matrix that transforms the left-handed
part of $u$-type ($d$-type) quarks in the bi-unitary diagonalization
of the mass matrices; namely, $V$ is unitary. Then, the orthogonality
relation of the $d$-column and $b$-column can be written as a
triangle relation:
\beq
     V_{ud} V^*_{ub} + V_{cd} V^*_{cb} + V_{td} V^*_{tb} = 0\,,\quad
  \raisebox{-0.5in}{\epsfig{file=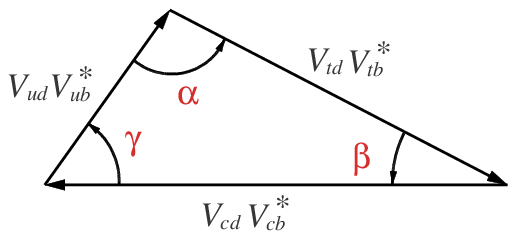,width=2in}}
\eeq
where the angles are defined by~\footnote{
Another common notation is $(\alpha,\beta,\gamma)\equiv(\phi_2,\phi_1,\phi_3)$.}
\beq
   \alpha \equiv
      \arg\left( {V_{td}V_{tb}^* \over -V_{ud}V_{ub}^*}\right),\;
   \beta  \equiv 
      \arg\left( {V_{cd}V_{cb}^* \over -V_{td}V_{tb}^*}\right),\;
   \gamma \equiv
      \arg\left( {V_{ud}V_{ub}^* \over -V_{cd}V_{cb}^*}\right) \,.
  \label{eq:angdef}
\eeq
Note that when quark phases are changed, the shape of the triangle is
invariant, and that if all $V_{ij}$ are real, the triangle reduces to a
line. Also, it should be emphasized that the sum of the angles is
always $\pi$ (mod $2\pi$) even if the triangle does not close. Thus,
$\alpha+\beta+\gamma = \pi$ does not test the unitarity; it simply
tests if the angles measured are as defined above. As long as
the test of unitarity is concerned, the measurements of the absolute
values of the sides of the triangle is just as important as those of
the angles.

Experimentally, the unitarity triangle is already over-constrained.
Primary inputs are, (1) $|V_{ub}/V_{cb}|$ from the semileptonic decays
of $B$, (2) $B^0$-$\bar B^0$ mixing which gives $|V_{td}|$,
(3) and $\epsilon_K$. The upper limit on the $B_s$ mixing also contribute, but
to a lesser degree than the above three. When the unitarity triangle is
normalized to the length of the bottom ($|V_{cd} V_{cb}|$), each of the
three measurements above form a band of for the location of the
tip of the triangle. Many such fit have been performed
and now the
consensus is that the three line cross at a single point within errors.
This is already supports the standard model of $CP$ violation.
In one such fit,\cite{Ciuchini+} the value of $\sin2\beta$ is predicted as
\beq
    \sin2\beta = 0.698 \pm 0.066\,.
  \label{eq:sin2bfit}
\eeq

$CP$ violation ($CPV$) in B decay may be classified into three categories:
\begin{enumerate}
 \item $CPV$ in the neutral $B$ mixing which manifests as the 
 particle-antiparticle imbalance in the physical neutral $B$ states ($B_{a,b}$);
 namely, $|\bra B^0 | B_{a,b}\ket |^2 \not= |\bra \bar B^0 | B_{a,b}\ket |^2$,
 \item  $CPV$ by the mixing-decay interference which can occur
 when both $B^0$ and $\bar B^0$ can decay to the same final state $f$, and
 \item $CPV$ in decay; namely, the asymmetries in instantaneous decay rates:
 $|Amp(B\to f)| \not= |Amp(\bar B\to \bar f)|$ which can happen when there
 are multiple diagrams with different weak phases and different strong
 phases.
\end{enumerate}

\section {$CPV$ in mixing}

Assuming $CPT$, the eigenstates of mass and decay rate can be
written as
\[
   \left\{
   \begin{array}{rclr}
      B_a &=& p B^0 + q \ovl B^0 & \quad (m_a, \gamma_a)\\
      B_b &=& p B^0 - q \ovl B^0 & (m_b, \gamma_b)
   \end{array} \right.\,.
\]
The asymmetry in $B^0,\bar B^0$ contents is the same
for $B_a$ and $B_b$ and given by
  \[
    \delta\equiv     
      {|\bra B^0|B_{a,b}\ket|^2 - |\bra\ovl B^0|B_{a,b}\ket|^2\over
       |\bra B^0|B_{a,b}\ket|^2 + |\bra\ovl B^0|B_{a,b}\ket|^2}
     = {|p|^2 - |q|^2 \over |p|^2 + |q|^2}\,.
  \]
The flavor contents may be measured by the lepton sign in the
semileptonic decays. Since $\gamma_a\sim\gamma_b$, one cannot
separate $B_a$ and $B_b$ by lifetime as in the case of 
the neutral kaon system. On $\Upsilon(4S)$, however, one can
measure the same-sign dilepton asymmetry where both $B$'s decay
semileptonically:\cite{Okun+}
  \[
     A_{\ell\ell}\equiv
    { N(\ell^+\ell^+) - N(\ell^-\ell^-) \over
      N(\ell^+\ell^+) + N(\ell^-\ell^-) } 
      \sim 2\delta\,.
  \]
There is also a $CP$ asymmetry in single lepton yield which can
be measured whenever equal number of $B$ and $\bar B$ are generated.\cite{Hagelin}
Assuming leptons from neutral and charged $B$'s cannot be separated,
  \[
    A_\ell\equiv 
     {N_{\Upsilon(4S)\to\ell^+} - N_{\Upsilon(4S)\to\ell^-} \over
      N_{\Upsilon(4S)\to\ell^+} + N_{\Upsilon(4S)\to\ell^-} } 
    = \chi\,\delta\,,\quad
    \chi (\hbox{mixing parameter})\sim 0.17\,.
  \]
This holds even for the quantum-correlated $B$ pair from
$\Upsilon(4S)$.\cite{HYlepton}

In the standard model, the dominant diagram for mixing is the box diagram
and gives
  \beq
   \raisebox{-0.5in}{\epsfig{file=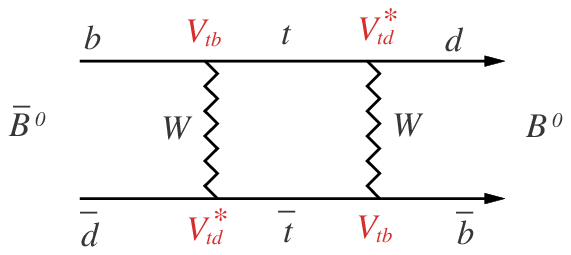, width=2in}} \quad
    {q\over p} = -{V^*_{tb} V_{td} \over V_{tb} V^*_{td}}\,
    \eta_B\,,
    \label{eq:pqB}
  \eeq
where $\eta_B$ is the (arbitrary) $CP$ phase of $B^0$:
$CP |B\ket = \eta_B |\bar B\ket$. We see that $q/p$ is a pure phase;
$|p| \not= |q|$ is caused at a higher-order by
the interference of the above diagram with the same one with 
$t$ replaced by $c$:
  \[ 
    \delta \sim -\,2\pi{m_c^2\over m_t^2}\,
    \Im\left( {V_{cb}V_{cd}^*\over V_{tb}V_{td}^*} \right)
    \; \sim 10^{-3}\;(\hbox{short distance})\,.
  \]
The value of $\delta$, however, is likely to be
dominated by long-distance effects such as 
$D\bar D$ intermediate states; it has a large theoretical
uncertainty and even the sign is not reliably predicted.\cite{Wolfen+}
This means that
one cannot determine CKM phases from $\delta$. 
If $\delta$ is found at percent level, however, it may signal
new physics, and its measurement has an engineering value 
since $\delta$ is assumed to be zero
in most calculations.

Experimental results are\cite{CLEOdelta,OPALdelta}
  \[
    \delta = \left\{
      \begin{array}{rcl}
         0.0070\pm0.0206\pm0.0030 & (CLEO 1993) \\
        -0.004\pm0.014\pm0.006    & (OPAL 1997) 
      \end{array} \right.\,,
  \]
where the OPAL result was actually obtained by 
fitting the time dependence of tagged
semileptonic decays of $B$'s on $Z^0$.

\section{$CPV$ by the mixing-decay interference}

  The flavor-tagged time-dependent decay distribution of the
  neutral $B$ meson system to a $CP$ eigenstate $f$~%
 \footnote{More precisely, we assumed $|q\ovl A/ p A|=1$.}
 is given by
  \beq
    \Gamma_{ B,\bar B
       \to f}(t) = N e^{-\gamma |t|} \left[
      1 \pm 
      \Im\left({q\ovl A\over pA}\right)
      \sin\delta m\,t \right] \,,
    \label{eq:upsdk}
  \eeq
  where $N$ is a normalization factor, $\delta m \equiv m_a - m_b$, 
  $A\equiv Amp(B^0\to f)$, $\bar A \equiv Amp(\ovl B^0\to f)$, 
  and we have assumed $\gamma_a = \gamma_b \equiv \gamma$.
  This expression applies not only to a pure $B^0$ or $\bar B^0$ state
  at $t=0$, but also to the $\Upsilon(4S)$ system by the replacement
  $t\to \Delta t \equiv t_1 - t_2$ where $t_1$ is the signal-side
  decay time and $t_2$ is the tagging-side decay time, and
  with the understanding that $\Gamma_{B\to f}$ ($\Gamma_{\bar B\to f}$)
  applies when the
  tagging-side was $\bar B^0$ ($B^0$). This is becuase, on $\Upsilon(4S)$,
  if one side decays as $B^0$ at a proper time $t_0$, then the other side
  is purely $\bar B^0$ at the same preper time $t_0$ and the evolution
  after that is the same as that of a single pure $\bar B^0$ prepared at
  time $t_0$. From (\ref{eq:upsdk}), the time-dependent asymmetry is simply,
  \beq
     A_{CP}(t) = \Im\left({q\ovl A\over pA}\right) \sin\delta m\,t \,.
   \label{eq:acpt}
  \eeq

\subsection{The gold-plated mode $J/\Psi K_S$}

  We can estimate $ \Im(q\ovl A / pA)$ for this mode as follows:
  Since the decay $\bar B^0 \to J/\Psi K_S$ is caused by the quark transition
  $b\to c \bar c s$, $\bar A$ contains 
  the CKM factor $V_{cb} V_{cs}^*$, and since $\bar K^0$ is observed as
  $K_S$, $\bar A$ should contain $\bra Ks|\bar K\ket$. Similarly,
  $\bar A$ contains $V_{cb}^* V_{cs}$ and $\bra Ks|K\ket$. In addition,
  when a state $|a\ket$ is related to its $CP$ conjugate state, there appears the
  $CP$ phase $\eta_a$ of that state: $CP|a\ket = \eta_a |\bar a\ket$. In particular,
  \[
      CP |\Psi K^0\ket = (-)^{L_{\Psi K}} \eta_\Psi\eta_K |\Psi \bar K^0\ket\,,
  \]
  where $L_{\Psi K}$ is the orbital angular momentum between $\Psi$ and $K$.
  Using the definition $K_S = p_K K^0 - q_K \bar K^0$ and the same procedure as
  (\ref{eq:pqB}),
  \[
     {\bra Ks|\bar K\ket \over \bra Ks|K\ket} = {-q_K^*\over p_K^*} =
     {V_{cd}^* V_{cs} \over V_{cd} V_{cs}^*}\eta_K^*\,.
  \]
  Then, the amplitude ratio $\bar A/A$ becomes
  \[
    {\bar A\over A} = \dspl{\bra Ks|\bar K\ket \over \bra Ks|K\ket}\,
         {\bra\Psi \bar K|H|\bar B\ket \over \bra\Psi K|H|B\ket}
      = \left[
      {V_{cd}^* V_{cs} \over V_{cd} V_{cs}^*}\,
       \eta_K^*  \right]\,
     \left[(-)^{L_{\Psi K}} \eta_\Psi\eta_K\,
       {V_{cb} V_{cs}^* \over V_{cb}^* V_{cs}} 
          \,\eta_B^* \right]  
  \]
  Combining this with  (\ref{eq:pqB}) and noting
  $\eta_\Psi = +1$ (regardless of the $CP$ phase of charm quark) 
  and $L_{\Psi K} = 1$, we get
  \[
     {q\ovl A\over pA} = 
      \left( {V_{cd} V_{cb}^*\over -V_{td}V_{tb}^*}\right)^*
      \Bigg/\left({V_{cd} V_{cb}^*\over -V_{td}V_{tb}^*}\right) \;\to\;
     \Im\left({q\ovl A\over pA}\right) = -\sin 2\beta
  \]
  where we have used the exact definition of the angle $\beta$ as given by
  (\ref{eq:angdef}).
  The arbitrary $CP$ phases $\eta_B$ and $\eta_K$ are canceled out, and
  the result is also invariant of the quark phases.

\begin{figure}
   \begin{center}
    \epsfig{file=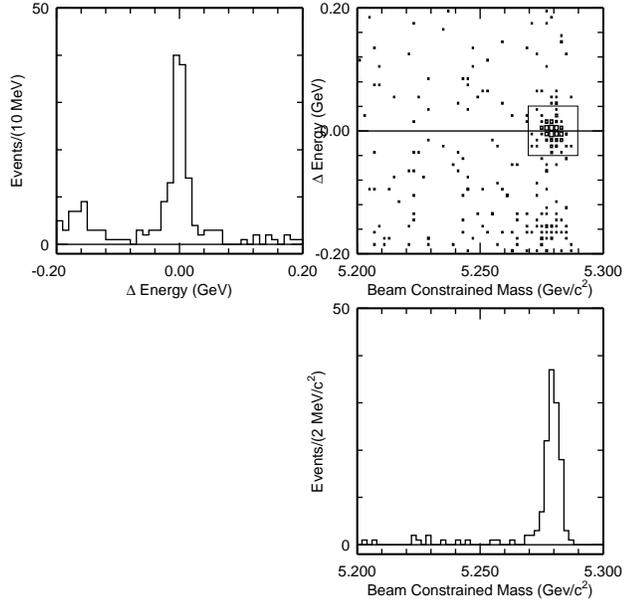,width=3.2in}
    \caption{The $\Delta E$-$m_{\rm bc}$ plot and 
     its projections for the $B\to J/\Psi K_S$ candidates (Belle).}
     \label{fg:psiks}
   \end{center}
  \end{figure}

\begin{figure}
   \begin{center}
    \epsfig{file=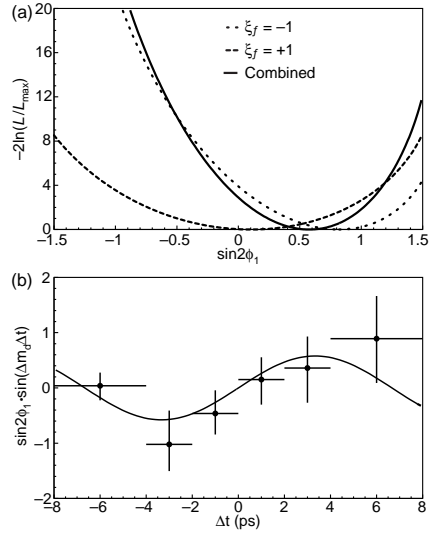,width=2.2in}
    \caption{(a) The $\chi^2$ plot for $sin2\beta(\sin2\phi_1)$ for $CP+$ modes,
      $CP-$ modes, and combined. (b) The time-dependent asymmetry adjusted 
      for $CP$ eigenvalues of the final state. (Belle)}
     \label{fg:sin2b}
   \end{center}
  \end{figure}

  Several experiments have attempted the measurement of $\sin2\beta$. Here,
  the analysis by Belle is shown because of the author's familiality with
  the experiment. In the c.m. system of $\Upsilon(4S)$, a $B$ meson has
  a fixed energy and a fixed absolute momentum. If it decays to a set of daughter
  particles, then
  \[
      E_{\rm tot} \equiv  \sum_{i=1}^n E_i =
     {m_{\Upsilon(4S)}\over 2} = 5.29 {\rm GeV}\,,\quad
      P_{\rm tot} = | \sum_{i=1}^n \vec P_i | = 0.34 {\rm GeV/c}\,,
  \]
  where $(E_i,\vec P_i)$ is the 4-momentum of the $i$-th daughter. One could
  thus plot $E_{\rm tot}$ vs $P_{\rm tot}$ to look for a peak at the 
  expected location; historically, however, two equivalent parameters,
  $\Delta E$ and $m_{\rm bc}$ (the `beam-constrained' mass) are used:
  \[
      \Delta E \equiv E_{\rm tot} - {m_{\Upsilon(4S)}\over 2}\,,\quad
     m_{\rm bc} \equiv 
     \sqrt{ \Big({m_{\Upsilon(4S)}\over 2}\Big)^2 - P_{\rm tot}^2 }\,.
  \]
  Figure~\ref{fg:psiks} shows the $\Delta E$-$m_{\rm bc}$ plot and 
  its projections for the $B\to J/\Psi K_S$ candidates
  corresponding to 10.5 fb$^{-1}$ of data.
 The analysis also used the modes
 $\Psi' K_S$, $\chi_{c1} K_S$, $\eta_c K_S$ ($CP=-1$)
 and $\Psi \pi^0$, $\Psi K_L$ $(CP=+1)$. The flavor tagging used
 $K^\pm$ and $\pi^\pm$ as well as leptons. The asymmetry flips sign
 for different $CP$ eigenvalues. The resulting value of $\sin2\beta(\sin2\phi_1)$
 and the time-dependent asymmetry is shown in Figure~\ref{fg:sin2b}.
  Figure~\ref{fg:psiks} shows the $\Delta E$-$m_{\rm bc}$ plot and 
  its projections for the $B\to J/\Psi K_S$ candidates.
 The measurements of $\sin2\beta$ are summarized in Table~\ref{tb:sin2b}.
 The numbers are consistent with the `prediction' (\ref{eq:sin2bfit}) of
 the standard model.
 \begin{table}
  \begin{center}
  \caption{Measurements of $\sin2\beta$.}
  \begin{tabular}{|c|@{\quad\tstrut{8}{4}}cl|}
    \hline
    experiment & $\sin2\beta$ & ref. \\ \hline
    OPAL  & $3.2^{+1.8}_{-2.0}\pm0.5$ & \cite{OPAL-sin2b} \\
    ALEPH & $0.84^{+0.82}_{-1.04}\pm0.16$ & \cite{Aleph-sin2b} \\
    CDF   & $0.79^{+0.41}_{-0.44}$ &  \cite{CDF-sin2b}    \\ 
    BaBar & $0.34\pm0.20\pm 0.05$ & \cite{BaBar-sin2b}\\
    Belle & $0.058^{+0.32}_{-0.34}{}^{+0.09}_{-0.10}$  & \cite{Belle-sin2b} \\
    \hline
  \end{tabular}
  \label{tb:sin2b}
  \end{center}
 \end{table}

\subsection{The $\pi^+\pi^-$ final state}

The tree diagram for $\bar B^0 \to \pi^+\pi^-$ is caused by the
quark-level transition $b\to u \bar u d$, and thus $\bar A/ A$ is
\[
   \raisebox{-0.3in}{ \epsfig{file=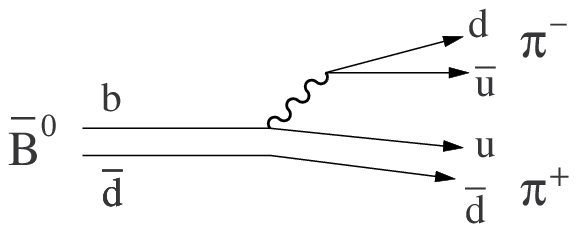, width=1.5in} }\qquad
   {\bar A\over A} =  {V_{ub} V_{ud}^*\over V_{ub}^* V_{ud}}\eta_B^*\,.
\]
Together with (\ref{eq:pqB}), the asymmetry coefficient for this mode is
\beqa
  \Im\left({q\over p}\cdot{\bar A\over A} \right) &=& \Im
       \left(-{V_{tb}^*V_{td}\over V_{tb}V_{td}^*}\eta_B \cdot
    {V_{ub} V_{ud}^*\over V_{ub}^* V_{ud}}\eta_B^*\right) \\
   &=& \Im \left[ -\left({V_{tb}^*V_{td}\over -V_{ub}V_{ud}^*}\right)\Bigg/
        \left({V_{tb}^*V_{td}\over -V_{ub}V_{ud}^*}\right)^* \right]
   = -\sin2\alpha\,,
\eeqa
where the arbitrary $CP$ phase $\eta_B$ again cancelled out and we have used
the definition of the angle $\alpha$ given in (\ref{eq:angdef}).

Since the $\pi^+\pi^-$ mode is already observed at $\sim1.5$ events/fb$^{-1}$,
we can expect about 450 events at 300 fb$^{-1}$ where the background would
have improved, say, by a better vertexing. This together with the effective
tagging efficiency of 0.27, the error on $\sin2\alpha$ will be about 0.15.
There is, however, a complication caused by the $b\to d$ penguin transition
which has a different weak phase from that of the tree transition.
Since the isospin-2 component does not receive
contribution from the penguin, one may extract it by combining with
$B^-\to\pi^-\pi^0$ and $\bar B^0\to\pi^0\pi^0$
and applying an isospin analysis.\cite{Gronau-London}
The detection of $\pi^0\pi^0$ mode, however, is experimentally
challenging and the method suffers from a reduction of statistical
power. A more promissing way may be provided by the QCD factorization
approach\cite{QCDfact} with a systematic hearvy-quark expansion which
indicates that $\sin2\alpha$ can be determined with a small
theoretical error (of order 0.1) from the asymmetry coefficient
$\Im(q\bar A/p A)$ albeit with a discrete ambiguity.

\subsection{Flavor-specific final states}

$CPV$ by mixing-decay inteference can occur even if the final state
is not a $CP$ eigen state as long as both $B^0$ and $\bar B^0$ can
decay to the same final state. One example is 
the $D^+\pi^-$ final state\cite{Dstpi}
where $\bar B^0\to D^+\pi^-$ caused by $b\to c\bar ud$ and 
$B^0\to D^+\pi^-$ cause by $\bar b\to \bar u c \bar d$ interfere through
mixing. The rate of a pure $\bar B^0$ at $t=0$ decaying to $D^+\pi^-$
at $t$ is
\[
   \Gamma_{\bar B^0\to D^+\pi^-}(t) \propto
      {e^{-\gamma |t|}\over2} \left[ (1+r^2) + (1-r^2)\cos\delta mt 
         +  2r\sin(\phi_w+\delta) \,\sin\delta mt \right] 
\]
with $r\equiv|A(B^0\to D^+\pi^-)/A(\bar B^0\to D^+\pi^-)\sim0.02$,
$\phi_w = 2\beta + \gamma$ according to the exact definitions
(\ref{eq:angdef})\cite{ASY} and  $\delta$ is the strong phase.
Starting from $\Gamma_{\bar B^0\to D^+\pi^-}$ given above, 
$\Gamma_{B^0\to D^+\pi^-}$ is obtained by flipping the signs
of $\cos\delta mt$ and $\sin\delta mt$, $\Gamma_{B^0\to D^-\pi^+}$
by $\phi_w\to -\phi_w$, and $\Gamma_{\bar B^0\to D^-\pi^+}$ by both
replacements. On $\Upsilon(4S)$, the only modification needed is
again $t\to \Delta t$.

If we set $\delta=0$, the $CP$ asymmetry between the two
favored modes ($\Gamma_{\bar B^0\to D^+\pi^-}$ and $\Gamma_{B^0\to D^-\pi^+}$)
is $\sim0.01\sin\phi_w$ and that between the two suppressed modes
 ($\Gamma_{\bar B^0\to D^-\pi^+}$ and $\Gamma_{B^0\to D^+\pi^-}$)
is $\sim0.06\sin\phi_w$. The statistics of the favored modes is about
5 times that of the suppressed modes; thus, the $CPV$ information is
mostly contained in the suppressed modes.
The measurement of these 4 modes give two quantities:
$r\sin(\phi_w-\delta)$ and $r\sin(\phi_w+\delta)$. Thus, the value
of $r$ needs to be input externally in order to extract 
$\phi_w = 2\beta + \gamma$.

One could also use $D^{*+}\pi^-$ where $D^0$ is not reconstructed
in the decay $D^{*+}\to D^0\pi^+$, which enhances the statistics.
The expected precision for a given luminosity may be expressed as
$\sigma_{\sin(2\beta+\gamma)} = 4\sim5 \sigma_{\sin2\beta}$.

A similar mechanism for $CPV$ can be found 
in the $D^{*+}\rho^-$ mode.\cite{ASY,LSS}
The asymmetries again are of order $1\sim5$\%; this time, however, there
are interferences among the three polarization amplitudes each of which
evolves as a function of time. One thus measures the angular
correlation of the decays $D^{*+}\to D^0\pi^+$ and $\rho^-\to\pi^+\pi^0$
at a given time. The relevant weak phase is the same as that of
$D^+\pi^-$: $\phi_w = 2\beta+\gamma$, but there are more degrees of
freedom for the measurements. The statistic-enhancing partial
reconstruction technique as the one used for $D^{*+}\pi^-$ is probably
not realistic due to the requirement to measure the decay angles.
The statistical power, however, is extected to be comparable to that
of $D^{*+}\pi^-$.

\section{$CPV$ in decay}

The particle-antiparticle asymmetry in partial decay rate can occur
when there are multiple diagrams with different weak phases (i.e. the
$CKM$ phases) and different strong phases. Here, we will look at
two historically important categories of modes: $DK$ and $K\pi$, $\pi\pi$
modes. There are, however, many other modes that are just as important in
studying $CP$ violation such as $B\to$ a light pseudoscalar plus a 
light vector.

\subsection{$B\to DK$}

One clean example is $B^- \to D_{1,2} K^-$ (and its charge
conjugate mode) where $D_{1,2}\equiv(D^0\pm\bar D^0)/\sqrt2$, $D_1$
is detected by the final states $K^+K^-$, $\pi^+\pi^-$, etc.
and $D_2$ by $K_S\pi^0$, $K_S \rho^0$, $K_S \phi$, etc.
Then, $A(B^- \to D_1 K^-)$ is the sum of 
$A(B^-\to D^0 K^-) \equiv a e^{i\phi_c} e^{\delta_c}$ and 
$A(B^-\to \bar D^0 K^-) \equiv b e^{i\phi_u} e^{i\delta_u}$,
where $\phi_{c,u}$ are the phases of the $CKM$ factors (`weak'
phases), $\delta_{c,u}$ are the strong phases, and $a,b$ are positive.
For the charge-conjugate modes, the $CKM$ factors are complex-conjugated
but the strong phases stay the same.
The decay rates of $B^\mp\to D_1 K^\mp$ are then
\beq
  \Gamma(B^\mp\to D_l K^\mp) = {|a|^2\over2}
    \big[1+r^2+(-)^l\, 2r\, \Re (e^{\pm i\Delta\phi}e^{i\Delta\delta}) \big]
\eeq
where $l=1,2$, $r\equiv b/a$, $\Delta\phi\equiv\phi_u-\phi_c$, and 
$\Delta\delta\equiv\delta_u-\delta_c$. For a given $l$, we see that
there is an particle-antiparticle asymmetry if $\Delta\phi\not=0$ {\it and}
$\Delta\delta\not=0$. 

Once $r$ is measured by flavor-specific modes of $D^0$,
the measurements of the two modes $\Gamma(B^\mp\to D_1 K^\mp)$ (or $l=$2) allows
a determination of $\Delta\phi$ and $\Delta\delta$ by a triangle 
construction.\cite{DKtriangle} Experimentally, however, it would be
simpler to fit simultaneously 
the all 4 numbers $\Gamma(B^\mp\to D_l K^\mp)$ $(l=1,2)$
(each normalized to $\Gamma(B^-\to D^0 K^-)$). There is an experimental
difficulty in measuring $r$ by hadronic final states because of
the doubly-cabbibo-suppressed decays\cite{ADS} which causes interference
between $A(B^-\to D^0 K^-)$ and $A(B^-\to \bar D^0 K^-)$.\cite{ADS}
This, however, can be used to extract $r$ as well as $\Delta\phi$ and 
$\Delta\delta$ by measuring at least modes, say, $B^-\to (K^+\pi^-)K^-$ and
$B^-\to (K^-K^+)K^-$ together with their charge-conjugate modes.\cite{ADS}

The relevant quark diagrams
can be better understood by systematically writing down
all diagrams for $B^-/\bar B^0 \to DK/D_s\pi$. For $B^-$, we have
\[
  \raisebox{-0.9in}{\epsfig{file=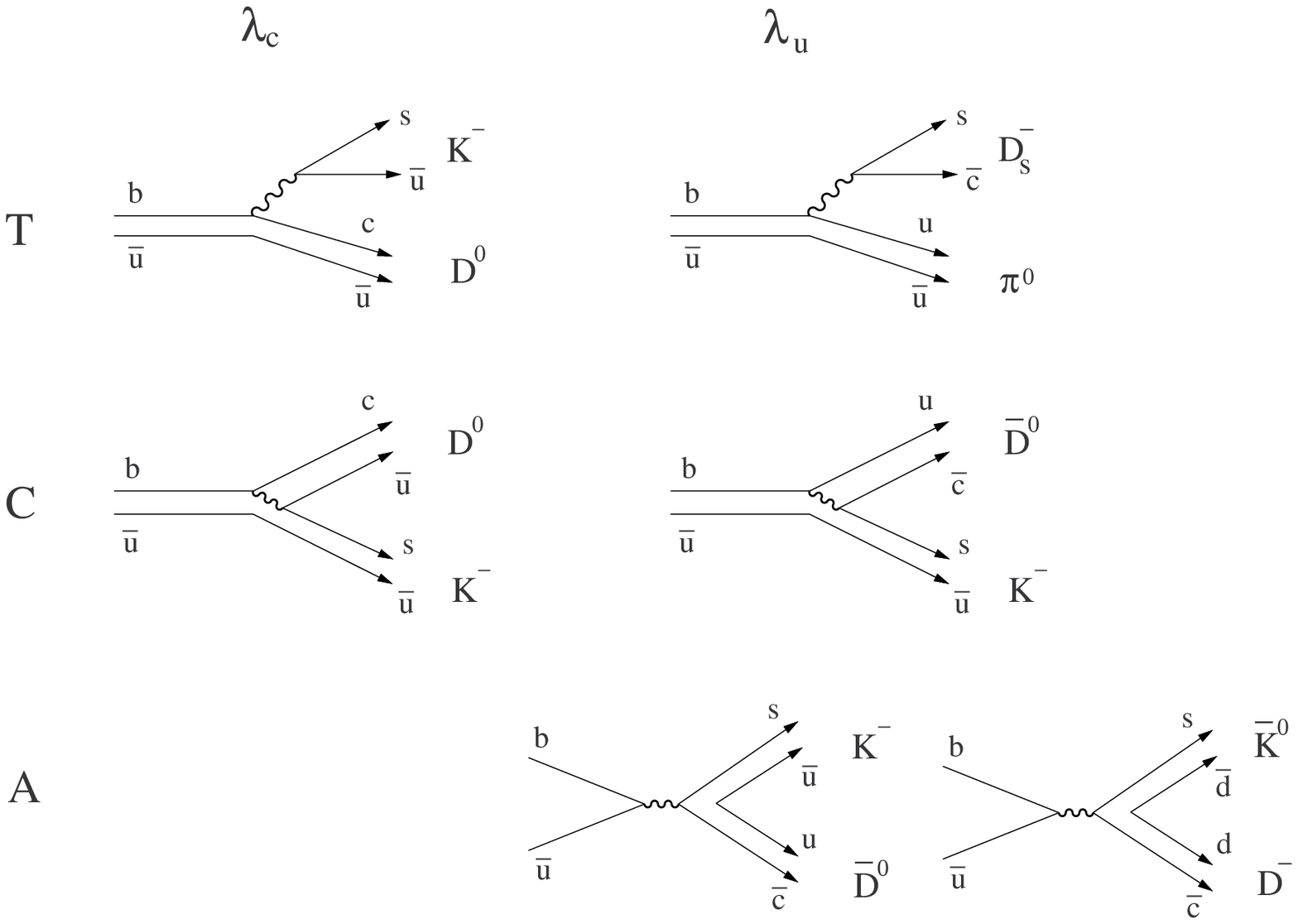,width=2.6in}}
  \begin{array}{ll}
    A(B^-\to D^0 K^- )  & =\lambda_c (T_c + C_c)  \\
    A(B^-\to \bar D^0 K^-) & =\lambda_u (C_u + A)   \\
    A(B^-\to D^- \bar K^0) & =\lambda_u A  \\
    A(B^-\to D_s^-\pi^0)   & ={1\over\sqrt2}\lambda_u T_u 
  \end{array}
\]
and those for $\bar B^0$ are 
\beq
  \raisebox{-0.7in}{\epsfig{file=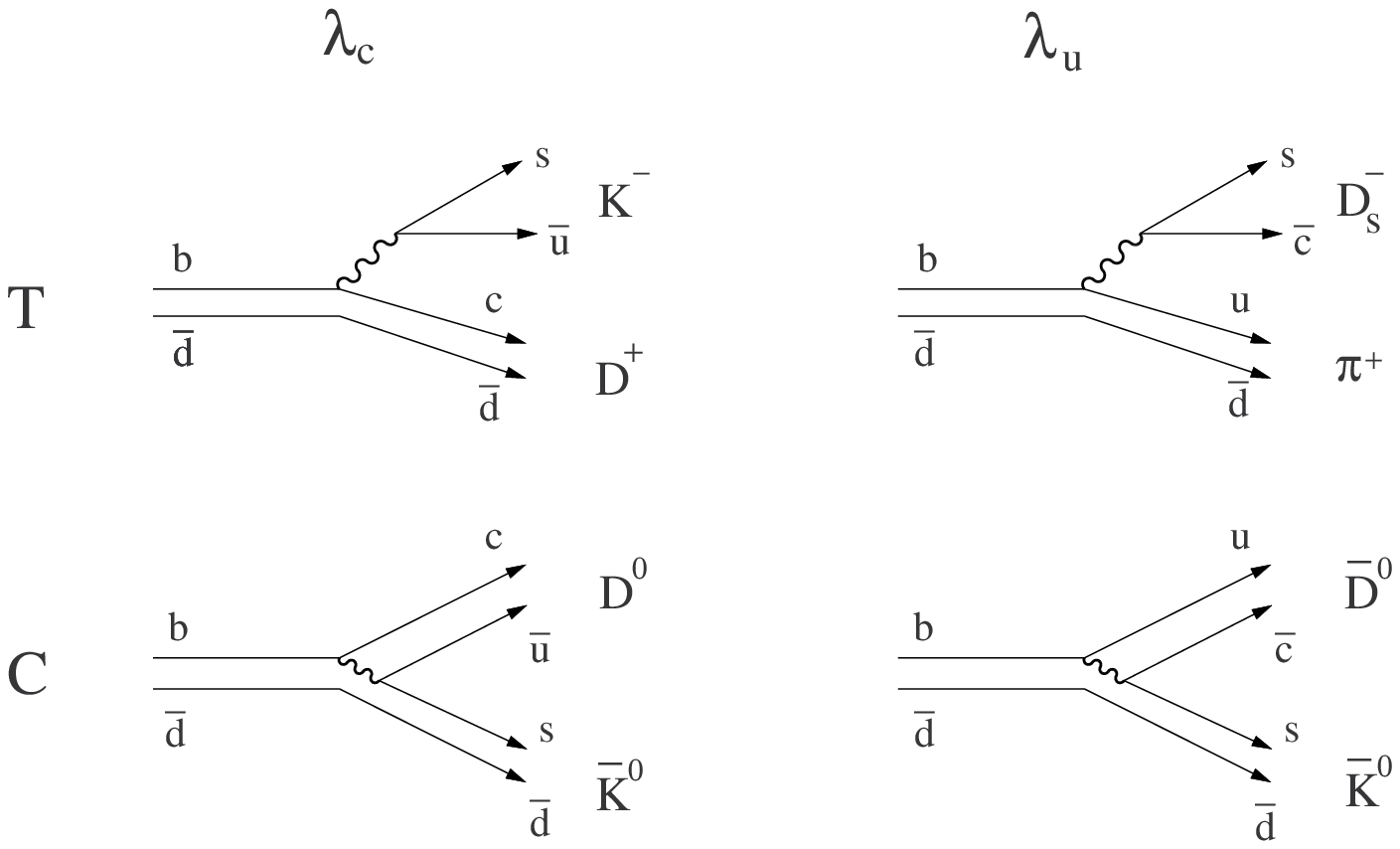,width=2.2in}}\qquad
  \begin{array}{ll}
    A(\bar B^0\to D^+ K^- )  & =\lambda_c T_c \\
    A(\bar B^0\to D^0 \bar K^0) & =\lambda_c C_c   \\
    A(\bar B^0\to \bar D^0 \bar K^0) & =\lambda_u C_u  \\
    A(\bar B^0\to D_s^-\pi^+)   & =\lambda_u T_u
  \end{array}
  \label{eq:B0Adef}
\eeq
where $\lambda_c\equiv V_{cb}V_{us}^*$ and $\lambda_u\equiv V_{ub}V_{cs}^*$ are
the $CKM$ factors. The strong phases are contained in 
$T_{c,u}$, $C_{c,u}$, and $A$, which are the tree, color-suppressed, and annihilation
amplitudes, respectively. One notes several interesting features: 
1. There is no penguin diagrams. This is because there should be even
number of $c$ or $\bar c$ quarks in the final state of a penguin diagram and
here we have one.
2. There is no annihilation diagram for $\bar B^0$. Such diagram should
have even number of $s$ or $\bar s$ quarks in the final state where we have
one.
3. We can read off the relations
\beqa
   A(D^0 K^-) &=& A(D^+ K^- ) + A(D^0 \bar K^0) \\
   A(\bar D^0 K^-) &=& A(D^- \bar K^0) + A(\bar D^0 \bar K^0)
   \label{eq:isoann}
\eeqa
which are nothing but the isospin relations and are valid even with
final-state rescatterings such as $B^-\to D_s^-\pi^0 \to \bar D^0 K^-$ 
or $D^-\bar K^0$. In fact, one can {\it define} $T_{c,u}$ and $C_{c,u}$
by (\ref{eq:B0Adef}) and $A$ by $A(D^- \bar K^0)$.
4. $D^- \bar K^0$ is a pure annihilation (including the rescattering).
If $A(D^- \bar K^0)$ turns out to be zero, one can extract $b$ from other
less-suppressed  modes.\cite{JangKo}

The measured weak phase is $\Delta\phi = \arg(\lambda_u/\lambda_c)\sim -\gamma$.
Strictly speaking, however, what is measured depends on the final state of the $D$
decay
\beq
   \Delta\phi = 
   \left\{
    \begin{array}{cr}
      -\gamma + \xi & \quad(K^+ K^-) \\
      -\gamma - \xi & (\pi^+\pi^-)
    \end{array}\right.\quad
    \xi\equiv\arg{V_{cd}V_{cs}^*\over -V_{ud}V_{us}^*}\,,
\eeq
where $\xi\sim\lambda^4 \sim 0.002$ in the standard model ($\lambda\sim0.22$ is
the Cabbibo suppression factor).
This difference is caused
by the small direct $CP$ violation in the $D$ decays.
Statistically, one needs about
300 fb$^{-1}$ or more for a viable measurement, and the suppression of background
for the suppressed modes is an experimental challenge.

\subsection{$B\to K\pi, \pi\pi$}

Tree-penguin interference could cause sizable rate asymmetries in these
modes.  Since many of them have already been observed, 
we may find rate  asymmetries sometime soon. The 
extraction of the angle $\gamma$, however, is non-trivial.
Ignoring the annihilation and electro-weak penguin processes,
the amplitudes for $\pi^-\pi^0$, $\bar K^0\pi^- $, and $K^-\pi^0$ can
be written as
\beq
  \raisebox{-0.6in}{\epsfig{file=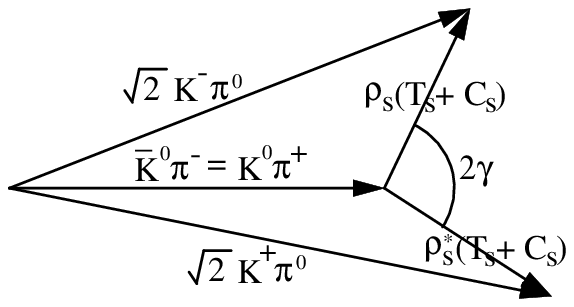,width=2.1in}}
  \begin{array}{rcl}
    \sqrt2 A(\pi^-\pi^0) &=& \rho_d (T_d + C_d)\\
    A(\bar K^0\pi^-) &=& \rho_t P  \\
    \sqrt2 A(K^-\pi^0) &=& \rho_d (T_s + C_s) + \rho_t P
  \end{array}
  \label{eq:kpi}
\eeq
where $\rho_d\equiv V_{ub}V_{ud}^*$, $\rho_s\equiv V_{ub}V_{us}^*$,
$\rho_t\equiv V_{tb}V_{ts}^*$, and $T_{d,s}$, $C_{d,s}$, $P$ are
the tree, color-suppressed tree, and $b\to s$ penguin amplitudes,
respectively. The subscripts on $T$ and $C$ refer to the
down-type quark generated by a $W$, and have no distinction
in the flavor $SU(3)$ limit.
For the charge-conjugate modes, the $CKM$ factors
are complex-conjugated and the rest stays the same. Taking
$\rho_t P$ as real, this leads to a double-triangle relation 
and the angle $\gamma$
(or more precisely, $\arg(\rho_t/\rho_d)$) can be extracted as shown,
where the value of $|\rho_d (T_s + C_s)|$ is estimated from 
$\pi^-\pi^0$:\cite{GRL}
\beq
 |\rho_d (T_s + C_s)| \sim \sqrt2 {\rho_s f_K \over \rho_d f_\pi}|A(\pi^-\pi^0)|\,.
  \label{eq:su3}
\eeq
The ratio of the decay constants accounts for the known part of the
$SU(3)$ breaking effect: $(T_s+C_s)/(T_d+C_d)\sim f_K/f_\pi$.
Note also that these modes are all charged $B$ modes and thus
self-tagging; namely, all detected events can be fully utilized.

The neutral $B$ modes such as $K_S \pi^0$ and $K^+\pi^-$	also
are useful modes in determining $\gamma$. 
As long as we can assume that the amplitudes $A(\bar B^0\to K^+\pi^-)$
and $A(\bar B\to K^0\pi^0)$ are zero, which is a good approximation in
the standard model, one does not need flavor-tagging
in measuring the absolute values of these decay modes even though
the process involves $B^0$-$\bar B^0$ mixing. When the relative
rates from $B^0$ and $\bar B^0$ are non-trivial and need to be
measured, as in the case of the $\pi^0\pi^0$ final state, then
flavor tagging is necessary. In such case, the decay time
measurement is useful but not required.

The annihilation process leads to the replacement 
$\rho_t P\to \rho_t P + V_{ub}V_{us}^* A$ 
whenever $\rho_t P$ appears in (\ref{eq:kpi}), where $A$ is the
annihilation amplitude (apart from the $CKM$ factor).
This keeps the triangle
relations intact but changes the meaning of the angle measured
because the annihilation part has the $CKM$ angle different from
that of the penguin part.
There may also be sizable $SU(3)$ breaking effect not yet accounted for
in (\ref{eq:su3}), but the largest uncertainty arises from the
electro-weak penguin processes which violate the isospin symmetry.
It effectively results in a correction of order unity in (\ref{eq:su3}).
Theoretical uncertainties at this time are thus quite large. A hope
is shed by the aforementioned recent development which
allows systematic analyses of factorization in the framework of
QCD.\cite{QCDfact} The theoretical errors are still not small, but
at least the uncertainties can now be estimated systematically.

\end{document}